# Frustrated $S = 1/2$ Chains in One-Dimensional Correlated Metal $Ti_4MnBi_2$

X. Y. Li[1]*, A. Nocera[1], K. Foyevtsova[1], G. A. Sawatzky[1, 2], M. Oudah[1], N. Murai[3], M. Kofu[3], M. Matsuura[4], H. Tamatsukuri[3], and M. C. Aronson[1, 2]

[1]Stewart Blusson Quantum Matter Institute, The University of British Columbia; Vancouver BC, V6T 1Z4, Canada.

[2]Department of Physics and Astronomy, The University of British Columbia; Vancouver BC, V6T 1Z1, Canada.

[3]J-PARC Center, Japan Atomic Energy Agency; Tokai, Ibaraki 319-1195, Japan.

[4]Comprehensive Research Organization for Science and Society; Tokai, Ibaraki 319-1106, Japan.

*Corresponding author. Email: xiyang.li@ubc.ca

**Electronic correlations lead to heavy quasiparticles in three-dimensional (3D) metals, and their collapse can destabilize magnetic moments. It is an open question whether there is an analogous instability in one-dimensional (1D) systems, unanswered due to the lack of metallic spin chain materials. We report neutron scattering measurements and Density Matrix Renormalization Group calculations establishing spinons in the correlated metal $Ti_4MnBi_2$, confirming that its magnetism is 1D. $Ti_4MnBi_2$ is inherently frustrated, forming near a quantum critical point (QCP) separating different temperature $T = 0$ phases of the $J_1$-$J_2$ XXZ model. 1D magnetism dominates to the lowest $T$, and is barely affected by weak interchain coupling. $Ti_4MnBi_2$ is the first metallic spin chain where 3D conduction electrons become strongly correlated due to their coupling to 1D magnetic moments.**

## Main

The stability of magnetic order in three-dimensional (3D) metals is driven by the interplay of magnetic interactions that can be frustrated, and proximity to electronic delocalization transitions such as Mott transitions or Kondo breakdown that destroy the magnetic moments themselves. Together these dual instabilities and their associated quantum fluctuations underlie a generic phase diagram for $T = 0$ magnetic order that has received substantial experimental support in *d*- and *f*-electron based metallic magnets[1,2].

Quantum fluctuations are particularly strong in one-dimensional (1D) systems, where powerful theory and the resilience of 1D character can be directly confronted in real materials[3]. Despite these considerable advantages, it has not yet been possible to integrate 1D magnetism into the phase diagram described above for 3D systems, given that there are only a few 1D systems that have itinerant states analogous to the conduction electrons in 3D metals. The organic conductors have no *d*- or *f*-electrons that could host localized moments, and instabilities of their 1D bands lead to spin density wave magnetism, unconventional superconductivity (SC), and metal-insulator transitions[4]. In the opposite limit, the localized *f*-electron moments in metallic $Yb_2Pt_2Pb$ display spinon excitations that are the signature of 1D magnetism, however, the coupling of those excitations to 3D conduction electrons is vanishingly weak[5]. Is it possible that there is a continuum of behaviors that connects these two endpoints in metallic 1D magnets, featuring the transformation of localized moments into correlated bands, via 1D versions of the Kondo breakdown or Mott-like transitions that are familiar from 3D metallic magnets?

We present here experimental and theoretical evidence that $Ti_4MnBi_2$ is just such a system, with 1D magnetism coupled to metallic electrons that are strongly correlated. $Ti_4MnBi_2$ consists of well-separated chains of spin $S = 1/2$ moments[6], shown by Density Functional Theory (DFT) calculations to originate with Mn-Ti molecular orbitals (MO). Inelastic neutron scattering (INS) measurements observe spinons, the hallmark excitation of 1D spin chains, and by comparison to Density Matrix Renormalization Group (DMRG) calculations, we show that the 1D character of $Ti_4MnBi_2$ is well described by the frustrated $J_1$-$J_2$ 1D XXZ Hamiltonian. DMRG reveals a complex $T = 0$ phase diagram, with antiferromagnetic (AF) $Ti_4MnBi_2$ forming close to ferromagnetic (FM) and vector chiral (VC) instabilities where quantum fluctuations minimize the growth of long-ranged and long-lived correlations. The magnetic moments are nearly localized in $Ti_4MnBi_2$, however their coupling to the 3D conduction electrons results in strong correlations and enhanced density of states, identified in both DFT and specific heat experiments.

## Electronic Structure and the $S$ = 1/2 Magnetic Moments in $Ti_4MnBi_2$

The remarkable 1D properties of $Ti_4MnBi_2$ originate with its structure (Fig. 1a), which features chains of Mn atoms separated by 7.4208(3) Å[7,8]. The small intrachain spacing of 2.4930(1) Å of the Mn atoms in this metallic system would ordinarily lead to itinerant magnetism[9], so it is surprising that Curie-Weiss fits to the magnetic susceptibility $\chi(T)$ above 50 K (SI 4.3, Fig. S9) reveal that the two Mn atoms per unit cell both have spin $S = 1/2$, with the Weiss temperature $\theta_W$ = -13.3(2) K indicating an AF mean field. DFT calculations highlight the central role of the Mn $d_{xy}$ and $d_{x^2-y^2}$ orbitals, and their dominance of the non-magnetic projected density of Mn states (PDOS) at the Fermi level $E_F$ (Fig. 1b, top panel). Their correlated nature is reflected in the sharp peak at $E_F$, responsible for an electronic instability into a state where the Mn $d_{xy}$ and $d_{x^2-y^2}$ electrons form localized moments. The Mn $d_{xy}$ and $d_{x^2-y^2}$ states are excluded from the Fermi energy (Fig. 1b, bottom panel) in the FM state, which approximates the true $Ti_4MnBi_2$ magnetic ground state, characterized by strong correlation and quantum fluctuation effects. The DOS at $E_F$ remains substantial, due to the formation of strong covalent bonds that couple the itinerant electronic states to the localized Mn $d_{xy}$ and $d_{x^2-y^2}$ states. These itinerant states are strongly correlated, yielding a calculated Sommerfeld coefficient $\gamma_{DFT}$ = 50 mJ/mol-K$^2$ (SI 3) that is in excellent agreement with the value $\gamma$ = 57 mJ/mol-K$^2$ found in specific heat measurements[6].

While electron itinerancy along the chain involves all of the Mn and Ti 3$d$-orbitals (Figs. 1c-d), the strong hybridization between the Mn $d_{xy}$ and $d_{x^2-y^2}$ and the Ti $d_{x^2-z^2}$ orbitals, together with their unusual square antiprismatic coordination (Fig. 1a), prompts a description in terms of charge localized into MO[10]. These MOs (Figs. 1e-f) are centered between the Mn atoms, each accommodating a single electron in their ground state yielding $S$ = 1/2 per MO. Gradient-corrected Local Density Approximation (LDA) calculations find that it is energetically favorable for the $S$ = 1/2 moments of the two MOs per unit cell to be aligned in parallel, a consequence of the strong Hund's interaction associated with $d^5$ $Mn^{2+}$. Not only does this imply that the near neighbor exchange $J_1$ is FM, but also that it is isotropic, like the Hund's interaction itself. The absence of FM signatures in the magnetization and $\theta_W < 0$ in $Ti_4MnBi_2$ suggest that $J_1$ competes with an AF second neighbor coupling $J_2$, ascribed to direct exchange or perhaps even the Ruderman-Kittel-Kasuya-Yosida (RKKY) interaction.

## $Ti_4MnBi_2$ and the Frustrated $J_1$-$J_2$ 1D XXZ Model

$Ti_4MnBi_2$ is best described as a system of spin $S$ = 1/2 chains with competing FM and AF exchange interactions. As we will show, impressive agreement between INS measurements and DMRG computations confirms that $Ti_4MnBi_2$ is a realization of the 1D $S$ = 1/2 $J_1$-$J_2$ XXZ model,

$$H = J_1 \sum_n \left[S_n^z \cdot S_{n+1}^z + \varepsilon_1\left(S_n^x \cdot S_{n+1}^x + S_n^y \cdot S_{n+1}^y\right)\right] + J_2 \sum_n \left[S_n^z \cdot S_{n+2}^z + \varepsilon_2\left(S_n^x \cdot S_{n+2}^x + S_n^y \cdot S_{n+2}^y\right)\right] \quad (1)$$

where the $S_n^a$ ($a$ = x, y, z) are components of the spin operator $S$ = 1/2 on neighboring (n, n+1) and next nearest neighbor (n, n+2) sites of a 1D chain. Guided by DFT, we take $J_2$ to be AF ($J_2 > 0$), and $J_1$ to be FM ($J_1 < 0$) with $\varepsilon_1 = 1$, reflecting the inferred isotropic character of $J_1$. The magnetization anisotropy (SI 4.3) shows that the $S = 1/2$ moments in $Ti_4MnBi_2$ have a pronounced easy-axis character related to $J_2$ (Fig. 1g), so that $\varepsilon_2 < 1$. The competition between $J_1$ and $J_2$ is controlled by the parameters $\alpha = J_2/|J_1|$, and $\varepsilon_1$, $\varepsilon_2$. Phase diagrams generated by DMRG track the magnetic gap $\Delta$ and the ↑↑↓↓ AF order parameter for the case of uniaxial anisotropy appropriate for $Ti_4MnBi_2$ (Fig. 1h). In the isotropic limit ($\varepsilon_1 = \varepsilon_2 = 1$), which is most appropriate for the oxide-based insulators studied so far (Table S3, Fig. S22), there is good agreement with previous studies [11–13], finding a gapless FM phase for $\alpha_C < 0.25$, and a gapped VC phase for $0.25 < \alpha_C \lesssim 0.4$ (SI 4.1). For uniaxial anisotropy ($\varepsilon_2 \rightarrow 0$), increasing $\alpha$ drives a transition from the gapless FM phase to a gapped phase with ↑↑↓↓ AF order (Fig. S5), as well as a previously reported FM phase with partial polarization[13–15]. Intermediate values of $\varepsilon_2$ lead to the collapse of the gapped ↑↑↓↓ phase, resulting in a VC phase with longer-ranged correlations and a vanishingly small gap that persists into the isotropic limit.

The detailed comparison of experiments and theoretical analysis presented here establishes that $Ti_4MnBi_2$ is the first metallic system that is well described by the frustrated $J_1$-$J_2$ 1D $S = 1/2$ XXZ Hamiltonian, and is also a rare example with pronounced easy-axis anisotropy. We will show that it forms very near the nexus of the FM, ↑↑↓↓, and VC states where the strongest QC fluctuations exist (Fig. 1h) [16–18].

## Elastic Scattering: Short-Ranged Magnetic Correlations

Long-range magnetic order is absent in $Ti_4MnBi_2$, although broad peaks are found near 2 K in the magnetic susceptibility $\chi(T)$ and the specific heat $C(T)/T$, suggesting that magnetic correlations are extremely short-ranged[6]. This expectation is confirmed by measurements of the elastic dynamical structure factor $M(\boldsymbol{Q})$ (Figs. 2a-c). A broad ridge of scattering is observed at 0.3

K that is centered at $\boldsymbol{Q}^*$ = 0.76(4) reciprocal lattice units (r.l.u.). Summing the elastic scattering over the transverse wave vectors $\boldsymbol{Q}_{\mathrm{HH}}$ reveals a broad peak in $M(\boldsymbol{Q})$ (Fig. 2d), with an intensity that increases with decreasing $T$, saturating at a value of $M_{coh}^2$ = 0.19(3) $\mu_B^2$/Mn below 2 K (Fig. 2f), as does a $\boldsymbol{Q}$-independent contribution $M_{inc}^2$ = 0.36(2) $\mu_B^2$/Mn (Fig. 2g). While there is a small reduction in the width of $M(\boldsymbol{Q})$ with decreasing $T$ (Fig. 2e), the spatial correlations associated with this peak never extend significantly beyond the unit cell. The growth of $M_{coh}^2$ and $M_{inc}^2$ as $T$ approaches ~ 2 K suggests the onset of weak magnetic order (Figs. 2f-g). The elastic scattering in $Ti_4MnBi_2$ is dominated by $M_{inc}^2$, which is $\boldsymbol{Q}_{\mathrm{L}}$ independent, and thus local in character (Fig. 2g).

## Inelastic Neutron Scattering: Spinons and Helimagnons

INS measurements of $M(\boldsymbol{Q}, E)$ reveal a broad continuum of excitations in $Ti_4MnBi_2$ that disperses along $\boldsymbol{Q}_{\mathrm{L}}$ (Fig. 3a), but not for transverse wave vectors $\boldsymbol{Q}_{\mathrm{HH}}$ (Fig. 3b). These excitations are consequently 1D, and are confined to the chains. A striking feature of the scattering along $\boldsymbol{Q}_{\mathrm{L}}$ is the extremely strong peak near $\boldsymbol{Q}_{\mathrm{L}}$ = 0, with a rapid dropoff in $\boldsymbol{Q}_{\mathrm{L}}$ that is primarily due to the magnetic form factor. Modelling of the form factor in $Ti_4MnBi_2$ (SI 4.5) reveals that the fluctuating moments are correlated over a length scale of ~ two unit cells along the chain axis, with a more gradual decrease in the transverse direction consistent with the $Mn^{2+}$ form factor.

DMRG computations capture the essential features of the INS spectrum as excitations of an underlying ↑↑↓↓ AF lattice within the $J_1$-$J_2$ XXZ model (Fig. 3c). This choice is consistent with the minima in the spectral dispersion occurring at $\boldsymbol{Q}_{\mathrm{L}}$ = 0, ±1, and not $\boldsymbol{Q}_{\mathrm{L}}$ = 0, 1, 2 as is found in the more familiar ↑↓↑↓ AF chain. The dispersions found in INS and DMRG in $Ti_4MnBi_2$ match best for the parameters $\boldsymbol{\alpha} = J_2/|J_1|$ = 0.75 (with $J_1$ = -2.8 meV and $J_2$ = 2.1 meV) and $\boldsymbol{\varepsilon}_2$ = 0.425 with a fixed value of $\boldsymbol{\varepsilon}_1$ = 1 (Figs. S16-S18). $Ti_4MnBi_2$ is located in the gapped ↑↑↓↓ phase, but very close to instabilities to the ungapped FM and VC phases (Fig. 1h). It has a pronounced easy-axis character with the transverse components dominating $M(\boldsymbol{Q}, E)$. The continua displayed by INS and DMRG are the analogs for frustrated $J_1$-$J_2$ chains[19] of the spinon continua that are the defining features of the Heisenberg and Ising AF $S = 1/2$ chains[20].

DMRG computations find that the magnetic excitations of the VC and ↑↑↓↓ phases are gapped over a broad range of $J_1$-$J_2$ model parameters (Fig. 1h), and the values of $\boldsymbol{\alpha}$, $\boldsymbol{\varepsilon}_2$ determined for $Ti_4MnBi_2$ give an excitation gap $\boldsymbol{\Delta} = 0.3$ meV (Fig. 3f). The energy dependencies of the structure factors $M(E)$ with $\boldsymbol{Q}_L = 0$ from INS and DMRG are compared in Fig. 3d, where the latter has been broadened from the INS instrumental resolution of 0.06 to 0.13 meV to match the INS data. High energy resolution DMRG calculations find a pronounced kink in $M(E)$ for $E = 0.35$ meV that marks the onset of the spinon spectrum at the gap edge (inset Fig. 3d). This feature is absent in the INS data and in the broadened DMRG results, presumably smeared beyond resolution. This excess broadening suggests that new physics is present in $Ti_4MnBi_2$ that is beyond the $J_1$-$J_2$ model.

DMRG finds a new branch of gapped excitations dispersing nearly linearly to $E = 0$ at $\boldsymbol{Q}^*_{\mathrm{DMRG}} = 0.70(2)$ r.l.u. (Fig. 3e). While there is no clear evidence for these excitations in the INS data, $\boldsymbol{Q}^*_{\mathrm{DMRG}}$ is very similar to $\boldsymbol{Q}^* = 0.76(4)$ r.l.u. of the broad elastic peak in $M(\boldsymbol{Q})$ (Fig. 2d). $Ti_4MnBi_2$ demonstrates the two periodicities expected for an AF helix, confirming that there is local VC character present in the gapped ↑↑↓↓ phase. The underlying AF lattice has ↑↑↓↓ order along the $c$-axis, leading to magnetic peaks at $\boldsymbol{Q}_{\mathrm{AF}} = (0, 0, \pm 1)$ r.l.u. that are not observed, since the moments are parallel to $\boldsymbol{Q}_{\mathrm{AF}}$. The precession of the moments in the $ab$ plane modulates this AF order along the $c$-axis, indicating that the broad elastic peak at $\boldsymbol{Q}^* = 0.76(4)$ r.l.u. is a satellite of the (0, 0, 1) magnetic peak with an incommensurate periodicity $1 - \boldsymbol{Q^*} = 0.24(4)$ r.l.u. that is close to four magnetic cells.

A holistic picture of this local VC phase comes from comparing the computed values of the excitation gap $\boldsymbol{\Delta}$, the AF order parameter $O_{\uparrow\uparrow\downarrow\downarrow}$, and the satellite wave vector $\boldsymbol{Q}^*$ of the helices as functions of $\boldsymbol{\alpha}$, with a fixed value of $\boldsymbol{\varepsilon}_2 = 0.425$ (Fig. 3f). All are zero for $\boldsymbol{\alpha} < \boldsymbol{\alpha}_C \approx 0.6$, consistent with this part of the $J_1$-$J_2$ XXZ phase diagram being a gapless FM. The onset of a gapped chiral phase coexisting with the ↑↑↓↓ phase for $\boldsymbol{\alpha} > \boldsymbol{\alpha}_C$ is evident from the steplike onset of the $O_{\uparrow\uparrow\downarrow\downarrow}$, in contrast to the more gradual increases of $\boldsymbol{\Delta}$ and $\boldsymbol{Q}^*$. The latter represents the helical modulation of the (0, 0, 1) AF Bragg peak, which becomes increasingly long wavelength as $\boldsymbol{Q}^* \rightarrow 1$ r.l.u. These observations suggest that $\boldsymbol{\alpha}_c = 0.6$, $\boldsymbol{\varepsilon}_2 = 0.425$ is a QCP that separates the ↑↑↓↓ AF phase from the FM and VC phases, analogous to the QCP for $\boldsymbol{\alpha}_C = 0.25$ reported in the isotropic limit. $Ti_4MnBi_2$ fortuitously forms very close to this QCP (Fig. 1h), where the spinon spectrum as well as the values of $\boldsymbol{Q}^*$, and the gap $\boldsymbol{\Delta}$ have their maximum sensitivities to the control parameters $\boldsymbol{\alpha}$ and $\boldsymbol{\varepsilon}_2$ (Fig. 3f).

Proximity to a QCP results in spatial and temporal correlations as $T \rightarrow 0$, and Figs. 4a-b provide an overview of the effects of temperature on $M(E)$. At 0.3 K, virtually all scattering is ascribed to the spinon continuum and to a resolution limited elastic peak. With increasing temperature, there is a decrease in the elastic scattering having energies less than 0.1 meV, and a matching increase in the spinon continuum, as mandated by the moment sum rule (Fig. 4c, SI 7). $M(E)$ grows dramatically for energies between the DNA instrumental resolution of 0.004 meV, and the onset of the nominally ungapped spinon spectrum above ~ 0.35 meV (Figs. 4a-b). Interestingly, $M(E)$ is proportional to the Bose factor (n+1) in this energy window, implying that the imaginary part of the dynamical susceptibility $\chi''(\boldsymbol{Q}, E) = \boldsymbol{\pi} M(\boldsymbol{Q}, E)/(n+1)$ is effectively energy independent and increases only weakly with temperature (inset Fig. 4a). It follows that the elastic peaks remain resolution limited and do not contribute to scattering, leading to the conclusion that there is no indication of critical dynamics at these energies and temperatures.

$\chi''(\boldsymbol{Q}, E)$ reveals (Fig. 4d) that the higher energy states are strongly impacted as the temperature is varied relative to the exchange interactions $J_1$ = -2.8 meV (~ 32 K) and $J_2$ = 2.1 meV (~ 24 K) that set the energy scale for the formation of the underlying ↑↑↓↓ AF lattice hosting the spinons. At the highest temperatures $k_BT >> J_1, J_2$, $\chi''(E)$ reveals a broad distribution of fluctuation energies, indicating paramagnetic (PM) fluctuations of the moment-bearing MOs. When $k_BT \sim J_1, J_2$, PM fluctuations subside and AF correlations begin to assemble into the underlying AF lattice, evident from the growing maximum in $\chi''(E)$ below ~ 25 K. $\chi''(E)$ increasingly resembles the $T$ = 0 DMRG spectrum, indicating that the spinon continuum in $Ti_4MnBi_2$ is fully formed and has become temperature-independent as $T$ approaches 2 K.

The Kramers-Kronig relation (Fig. 4e, SI 2.5) links the static susceptibility $\chi(T)$ to $\chi''(\boldsymbol{Q}, E, T)$. $\chi(T)$ displays a Curie-Weiss temperature dependence for temperatures above ~ 25 K, in good agreement with the values obtained from the Kramers-Kronig analysis. There is an increasing discrepancy between the two values at lower temperatures (Fig. 4e) that is due to the increased susceptibility associated with magnetic states having energies less than the experimental resolution of 0.1 meV that are not accounted for in this analysis. That missing low energy susceptibility grows from zero at 10 K to almost 60% of the total at 2 K, indirect evidence that the dynamical magnetic susceptibility associated with the slowest dynamics is increasingly enhanced as $T$ is reduced below 10 K. These dynamics, if present, have energies smaller than those accessed in the DNA and AMATERAS experiments.

## 1D Magnetism coupled to 3D Conduction Electrons

The potent combination of INS measurements and DMRG calculations has shown that $Ti_4MnBi_2$ is well described by the frustrated $J_1$-$J_2$ 1D XXZ Hamiltonian, where the low energy excitations are spinons that emerge from a 1D spin chain with ↑↑↓↓ AF order, with a predicted gap of ~ 0.35 meV due to its easy axis anisotropy. $Ti_4MnBi_2$ is located near gapless FM and VC phases, and this leads to the helical modulation of the ↑↑↓↓ AF order, evidenced by a nascent magnetic Bragg peak, accompanied by a branch of helimagnon excitations in the DMRG calculations.

Strong quantum fluctuations in 1D limit the development of spatial and temporal correlations that might otherwise culminate in order. Magnetic order requires interchain coupling to organize correlated regions in individual chains[21–23] where the confinement of the spinons results in a gap that cannot significantly exceed 0.35 meV in $Ti_4MnBi_2$. Even so, $Ti_4MnBi_2$ is only able to eke out the weakest of magnetic order at 2 K, involving only ~ 10% of the full $S = 1/2$ moment[6], and with spatial correlations that barely extend beyond the unit cell. Given the unusual weakness of its interchain coupling, the 1D quantum fluctuations remain unexpectedly strong in $Ti_4MnBi_2$ even at the lowest temperatures. In this way, $Ti_4MnBi_2$ is very close to fulfilling a necessary condition for realizing a 1D quantum spin liquid, which is the absence of magnetic order.

$Ti_4MnBi_2$ is only the second metallic system with 1D magnetism that has been reported, and we have shown that DFT calculations are able to differentiate between the localized *d*-electron states that bear the $S = 1/2$ moments, and the itinerant electrons that are coupled to them. The bands near the Fermi energy $E_F$ are complex and dispersing, suggesting the conduction electrons have dominantly 3D character. Accordingly, only weak anisotropy is found in the resistivity $\rho$ measured along the chains ($\rho_{001}$) and perpendicular ($\rho_{110}$) with $\rho_{110}/\rho_{001} \sim 2$ (Fig. S12). As well, Fermi Liquid (FL) properties $\rho(T) \sim AT^2$ ($A = 1.1\times10^{-3}$ μΩ-cm/K$^2$) and $C \sim \gamma T$ ($\gamma = 57$ mJ/mol-K$^2$) are observed in $Ti_4MnBi_2$ at low temperature[6], the latter in excellent agreement with the value $\gamma = 50$ mJ/mol-K$^2$, found here by DFT, among the largest values found in Mn-based metals[24]. The Kadowaki-Woods ratio $A/\gamma^2$ for $Ti_4MnBi_2$ is similar to those of conducting cuprates and ruthenates[25]. This attests to strong coupling between the 1D magnetic subsystem and the 3D conduction electrons in $Ti_4MnBi_2$ that is wholly absent in $Yb_2Pt_2Pb$.

Kondo physics provides direct evidence for coupling between the 1D moments and the 3D conduction electrons, and in the spirit of 3D Kondo lattices[26], potentially competes with magnetic order in correlated $Ti_4MnBi_2$. A Kondo insulator state is expected if the Kondo coupling $J_K \sim J_1$, $J_2$[27,28], and the absence of a large magnetic gap rules out this possibility in $Ti_4MnBi_2$. The conduction electrons retain their FL properties in $Ti_4MnBi_2$, although they are predicted to assume non-FL properties if the Kondo coupling to the 3D conduction electrons is sufficiently strong[29]. Finally, the magnitude of the moment in $Ti_4MnBi_2$ is robustly 1.73 $\mu_B$/Mn, indicating that the Kondo scale is at most emergent, akin to the ordering temperature $T_N$, or the magnetic gap $\Delta$. Kondo coupling first impacts the lowest energy states[27,28,30], and so it is possible that the broadening of the spinon spectrum by ~ 0.13 meV, substantially exceeding the instrumental resolution, may be an indication of weak Kondo coupling[31,32]. Significantly, this effect is not observed in insulating $KCuF_3$[33] or metallic $Yb_2Pt_2Pb$[5] where the Kondo effect is definitively absent. There is as well a roughly energy independent density of states present for the INS with $E < 0.1$ meV, potentially related to the initial stages of the formation of a quasielastic Kondo peak in $Ti_4MnBi_2$. While further experimental and theoretical work is required to clarify this situation, there is no definitive evidence at present for Kondo physics in $Ti_4MnBi_2$.

## Summary and Outlook

$Ti_4MnBi_2$ is not the first metallic spin chain system, but it is the first in which the coupling is sufficiently strong that the FL becomes significantly correlated, but not so strong that the 1D moments are suppressed. The intermetallic character of $Ti_4MnBi_2$ and related compounds[7,8] provides considerable flexibility in terms of compositional variation and chemical pressure that could be used to tune this 1D-3D coupling, and as well presents opportunities to extend these studies to different systems with different types of magnetic moments, and to conduction electrons with differing dimensionalities and types of spin and charge excitations. By analogy to the 3D heavy fermions, it is of great interest to seek new 1D metallic magnets with QCPs that might be accompanied by emergent instabilities such as unconventional SC and electronic delocalization, and as well new excitations leading to non-FL behavior. Given the weak interchain coupling in $Ti_4MnBi_2$, the 1D behavior is unusually persistent. It seems possible that these putative QCPs may be dominantly 1D, a consequence of $Ti_4MnBi_2$'s placement within the rich $T = 0$ phase diagram of the $J_1$-$J_2$ XXZ model. $Ti_4MnBi_2$ is the first system demonstrating an appreciable coupling between a 1D spin system and 3D conduction electrons, opening the door to the realization of a 1D Kondo effect and 1D heavy fermions in related systems that are yet to be discovered.

There remains a pressing need to discover new metallic 1D magnets where these compelling issues can be explored, and where a broader taxonomy of 1D metallic magnetism can be established that spans localized $Yb_2Pt_2Pb$, correlated $Ti_4MnBi_2$, and the organic conductors. By proving that this middle ground exists, $Ti_4MnBi_2$ is an important step towards establishing this broader landscape, so far little explored.

## References

1. Coleman, P. & Nevidomskyy, A. H. Frustration and the Kondo Effect in Heavy Fermion Materials. *J. Low Temp. Phys.* **161**, 182–202 (2010).

2. Si, Q. Quantum criticality and global phase diagram of magnetic heavy fermions. *Phys. status solidi B* **247**, 476–484 (2010).

3. Giamarchi, T. *Quantum Physics in One Dimension*. (Oxford University Press, 2003).

4. Jerome, D. & Bourbonnais, C. Quasi one-dimensional organic conductors: from Fröhlich conductivity and Peierls insulating state to magnetically-mediated superconductivity, a retrospective. *Comptes Rendus Phys.* **25**, 17–178 (2024).

5. Wu, L. S. *et al.* Orbital-exchange and fractional quantum number excitations in an *f*-electron metal, $Yb_2Pt_2Pb$. *Science* **352**, 1206–1210 (2016).

6. Pandey, A. *et al.* Correlations and incipient antiferromagnetic order within the linear Mn chains of metallic $Ti_4MnBi_2$. *Phys. Rev. B* **102**, 014406 (2020).

7. Richter, C. G., Jeitschko, W., Künnen, B. & Gerdes, M. H. The Ternary Titanium Transition Metal Bismuthides $Ti_4TBi_2$ with $T$ = Cr, Mn, Fe, Co, and Ni. *J. Solid State Chem.* **133**, 400–406 (1997).

8. Rytz, R. & Hoffmann, R. Chemical bonding in the ternary transition metal bismuthides $Ti_4TBi_2$ with $T$ = Cr, Mn, Fe, Co, and Ni. *Inorg. Chem.* **38**, 1609–1617 (1999).

9. Wada, H., Nakamura, H., Yoshimura, K., Shiga, M. & Nakamura, Y. Stability of Mn moments and spin fluctuations in $RMn_2$ (R: Rare earth). *J. Magn. Magn. Mater.* **70**, 134–136 (1987).

10. Jin, Z. *et al.* Magnetic molecular orbitals in MnSi. *Sci. Adv.* **9**, eadd5239 (2023).

11. Sirker, J. *et al.* $J_1$-$J_2$ Heisenberg model at and close to its $z = 4$ quantum critical point. *Phys. Rev. B* **84**, 144403 (2011).

12. Furukawa, S., Sato, M., Onoda, S. & Furusaki, A. Ground-state phase diagram of a spin-1/2 frustrated ferromagnetic XXZ chain: Haldane dimer phase and gapped/gapless chiral phases. *Phys. Rev. B* **86**, 094417 (2012).

13. Ueda, H. & Onoda, S. Roles of easy-plane and easy-axis XXZ anisotropy and bond alternation in a frustrated ferromagnetic spin-1/2 chain. *Phys. Rev. B* **101**, 224439 (2020).

14. Igarashi, J. I. Ground State and Excitation Spectrum of a Spin-1/2 Ising-Like Ferromagnetic Chain with Competing Interactions. *J. Phys. Soc. Japan* **58**, 4600–4609 (1989).

15. Tonegawa, T., Harada, I. & Igarashi, J. Ground-State Properties of the One-Dimensional Anisotropic Spin-1/2 Heisenberg Magnet with Competing Interactions. *Prog. Theor. Phys. Suppl.* **101**, 513–527 (1990).

16. Drechsler, S. L. *et al.* Frustrated cuprate route from antiferromagnetic to ferromagnetic spin-1/2 heisenberg chains: $Li_2ZrCuO_4$ as a missing link near the quantum critical point. *Phys. Rev. Lett.* **98**, 077202 (2007).

17. Sirker, J. Thermodynamics of multiferroic spin chains. *Phys. Rev. B* **81**, 014419 (2010).

18. Furukawa, S., Sato, M. & Onoda, S. Chiral Order and Electromagnetic Dynamics in One-Dimensional Multiferroic Cuprates. *Phys. Rev. Lett.* **105**, 257205 (2010).

19. Ren, J. & Sirker, J. Spinons and helimagnons in the frustrated Heisenberg chain. *Phys. Rev. B* **85**, 140410(R) (2012).

20. Tennant, D. A., Cowley, R. A., Nagler, S. E. & Tsvelik, A. M. Measurement of the spin-excitation continuum in one-dimensional $KCuF_3$ using neutron scattering. *Phys. Rev. B* **52**, 13368–13380 (1995).

21. Lake, B. *et al.* Confinement of fractional quantum number particles in a condensed-matter system. *Nat. Phys.* **6**, 50–55 (2010).

22. Gannon, W. J. *et al.* Spinon confinement and a sharp longitudinal mode in $Yb_2Pt_2Pb$ in magnetic fields. *Nat. Commun.* **10**, 1123 (2019).

23. Wu, L. S. *et al.* Tomonaga–Luttinger liquid behavior and spinon confinement in $YbAlO_3$. *Nat. Commun.* **10**, 698 (2019).

24. Li, X. Y. *et al.* Neutron scattering study of the kagome metal. *Phys. Rev. B* **104**, 134305 (2021).

25. Jacko, A. C., Fjærestad, J. O. & Powell, B. J. A unified explanation of the Kadowaki-Woods ratio in strongly correlated metals. *Nat. Phys.* **5**, 422–425 (2009).

26. Doniach, S. The Kondo lattice and weak antiferromagnetism. *Phys. B+C* **91B**, 231–234 (1977).

27. Tsunetsugu, H., Sigrist, M. & Ueda, K. The ground-state phase diagram of the one-dimensional Kondo lattice model. *Rev. Mod. Phys.* **69**, 809–863 (1997).

28. Nikolaenko, A. & Zhang, Y. H. Numerical signatures of ultra-local criticality in a one dimensional Kondo lattice model. *SciPost Phys.* **17**, 034 (2024).

29. Classen, L., Zaliznyak, I. & Tsvelik, A. M. Three-Dimensional Non-Fermi-Liquid Behavior from One-Dimensional Quantum Critical Local Moments. *Phys. Rev. Lett.* **120**, 156404 (2018).

30. Our initial DMRG calculations incorporating Kondo coupling between a 20 site XXZ spin chain and a 1D metallic bath indicate only subtle changes in the low energy magnetic spectrum for $J_K < 0.75 J_1, J_2$ (2024).

31. Laflorencie, N., Sørensen, E. S. & Affleck, I. The Kondo effect in spin chains. *J. Stat. Mech. Theory Exp.* **2008**, P02007 (2008).

32. Schimmel, D. H., Tsvelik, A. M. & Yevtushenko, O. M. Low energy properties of the Kondo chain in the RKKY regime. *New J. Phys.* **18**, 053004 (2016).

33. Scheie, A. *et al.* Erratum: Witnessing entanglement in quantum magnets using neutron scattering[Phys. Rev. B 103, 224434 (2021)], *Phys. Rev. B* **107**, 059902 (2023).

# Methods

## Sample growth and characterization

$Ti_4MnBi_2$ single crystals were grown using the flux method described in our previous work [6]. We optimized the growth conditions to prepare large single crystals for neutron scattering experiments. The crystals are rodlike, with typical dimensions of ~1 mm × 1 mm square cross-section and ~5-10 mm in length. X-ray diffraction experiments were carried on a powder prepared from single crystals out using a Bruker D8 Advance powder x-ray diffractometer. The crystals are single phase and the expected structure was confirmed [7,8] (Fig. S1a). The crystals have shiny metallic surfaces normal to the (110) and equivalent crystal directions and the (001) crystal direction is along the rod axis (Fig. S1b). The double-sided sample used for the inelastic neutron scattering (INS) experiments (AMATERAS and DNA@J-PARC) was assembled by co-aligning ~400 crystals on both sides of two 0.3 mm thick aluminum sheets, using hydrogen-free Cytop CTL-809M as the adhesive [34] (Fig. S1c). The sample size is roughly 20 mm (width) * 30 mm (height) * 4 mm (thickness) with a total mass of 10.2 g $Ti_4MnBi_2$ single crystals. The (110) axis of the crystals are normal to the aluminum sheets, and the scattering plane is ($H$, $H$, $L$). The neutron diffraction peaks in the ($H$, $H$, $L$) scattering plane are shown in Fig. S1d, and their sharpness as well as the relatively narrow rocking curve for the sample assembly (Fig. S1e) shows that the alignment of the crystals is excellent. The bulk magnetization was measured on a $Ti_4MnBi_2$ single crystal sample using a 7 T Quantum Design Magnetic Property Measurement System (MPMS3) magnetometer with temperatures ranging from 1.8 to 300 K, while a $^3$He insert extended those measurements to a base temperature of 0.4 K. The sample rotation measurements were performed at temperatures of 1.8, 5, 10, 25, and 100 K, with magnetic fields of 0.1, 1, 3, and 7 T, respectively. The rotation axis is in the horizontal plane, and the crystal was oriented with either the (001) or the (110) directions lying in that plane. The empty rotator was used as a background.

**Inelastic neutron scattering**

Our experiments were carried out at the AMATERAS [35] and DNA [36] instruments at the Materials and Life Science Experimental Facility (MLF) at J-PARC. The 10.2 g co-aligned sample (Fig. S1c) was mounted in the (*H*, *H*, *L*) scattering plane with the (1-10) direction vertical. Both experiments used the same $^3$He cryostat sample environment with a base temperature of 0.3 K. For measurements using the direct geometry instrument AMATERAS, the chopper configurations were set to select multiple incident energies $E_i$ of 3.13518, 7.73595, 15.1464, and 41.9667 meV with corresponding energy resolutions $\Delta E$ (full width at half maximum of the elastic peaks) of, respectively, 0.0581, 0.2244, 0.5652, and 2.4048 meV. The beam size was defined by slits to be 25 mm (width) * 35 mm (height), so that the sample with dimensions of 20 mm (width) * 30 mm (height) is fully illuminated by the neutron beam. The AMATERAS measurements were performed at 0.3, 1, 2, 5, 10, 25, and 100 K. The sample rotation angle is from -40˚ to 140˚ with a 0.5˚ increment at 0.3 K and a 1˚ increment at other temperatures. The data collecting time is ~30 hours at 0.3 K and ~12 hours at other temperatures. The initial data reduction was completed using the software suite UTSUSEMI [37]. The AMATERAS detectors are position sensitive along the vertical direction, which provides access to the out-of-plane (*H*, -*H*, *L*) and allows isolating contributions from the (*H*, *H*, *L*) scattering plane. The data obtained at 100 K is used to providing a temperature-independent background that was subtracted from the data measured at different temperatures. For the inverse geometry instrument DNA, the chopper configurations were set to high-resolution mode with $E_f$ = 2.08 meV with energy resolution of $\Delta E$ = 0.004 meV (full width at half maximum of the elastic peaks) which can measure -0.03 meV < $E$ < 0.1 meV range. The beam size was 20 mm (width) * 30 mm (height), well matched to the sample size. The measurements were performed at 0.3, 1, 1.4, 2, and 5 K, respectively. The sample rotation angle is from -40˚ to 140˚ with 1˚ increment, and the data collecting time is ~24 hours, except for the 5

K data, which were measured from -40° to -19° with 1° increment. Another high-flux mode scan was also performed at 0.3 K with the sample rotated from -10.2° to -4.2° with 1° increment, which provides data in the -0.5 meV < $E$ < 1.5 meV range. The data reduction and analysis were completed using the software suite UTSUSEMI [37] as well as Mslice/DAVE and PAN/DAVE [38]. For both AMATERAS and DNA data, the neutron absorption correction, including both in-plane and out-of-plane directions, was carried out using Mslice/DAVE [38], including the absorption cross section, as well as the coherent and incoherent scattering cross sections. The INS intensities were converted into absolute scattering cross sections by comparing to measurements of vanadium standards that were obtained in the same scattering conditions as used for the sample. More details are given in Supplementary Text 1-2.

**First-principles density-functional theory (DFT) calculations**

The DFT calculations of the electronic structure of $Ti_4MnBi_2$ were performed using the augmented plane-wave all-electron package WIEN2k [39] and the gradient-corrected local density approximation (GGA) by Perdew, Burke, and Ernzenhof [40] for the energy functional. The basis set size is set by choosing RKmax = 7.0, while the BZ integration is performed using a $\Gamma$-centered 10 × 10 × 10 $k$-vector grid. The considered structural model of $Ti_4MnBi_2$ is based on the experimentally determined tetragonal unit cell with the space group *I4/mcm* and the lattice vectors equal to $a = b = 10.547214$ Å and $c = 4.974468$ Å [7,8]. More details are given in Supplementary Information Section S.3.

**Density matrix renormalization group (DMRG) calculations**

We solved the Schrodinger equation with the $J_1$-$J_2$ Hamiltonian using the DMRG method [41,42] implemented in the DMRG++ code [43]. In the presence of the magnetic field, we computed the orientation-dependent magnetization as a function of the field angle, the spin-spin correlations, and the longitudinal $S_{ij}^{zz}(E)$ and transverse $S_{ij}^{+-}(E)$ spin structure factors in real space using the DMRG correction-vector method [44,45]. The corresponding correlation functions in momentum space were then obtained by a Fourier transform. When calculating the dynamical correlation functions, we fixed the broadening coefficient to $\eta = 0.06J_2$ and computed the spectral functions for each $E$ using the root-$N$ Correction-Vector algorithm with Krylov decomposition and a two-site DMRG update recently introduced [45], as implemented in the DMRG++ code [43]. We used $N$ = 8 and kept up to $m$ = 800 states. To avoid the necessity of reorthogonalizing the Krylov vectors, we allowed up to 200 Krylov vectors and truncated the effective Hamiltonian decomposition with a tolerance of $10^{-12}$. A representative transverse dynamic structure factor $S_{XX}(\boldsymbol{Q}, E)$ and longitudinal dynamic structure factor $S_{ZZ}(\boldsymbol{Q}, E)$ are shown in Fig. S8. More details are given in Supplementary Information Section S.4.

## Data availability

All data needed to evaluate the conclusions in the paper are available in the main text or the supplementary information. Raw neutron scattering data acquired in this study are preserved indefinitely at J-PARC.

## References

34. Rule, K. C., Mole, R. A. & Yu, D. Which glue to choose? A neutron scattering study of various adhesive materials and their effect on background scattering. *J. Appl. Crystallogr.* **51**, 1766–1772 (2018).

35. Nakajima, K. *et al.* AMATERAS: A Cold-Neutron Disk Chopper Spectrometer. *J. Phys. Soc. Japan* **80**, SB028 (2011).

36. Kawakita, Y. *et al.* Recent Progress on DNA ToF Backscattering Spectrometer in MLF, J-PARC. *EPJ Web Conf.* **272**, 02002 (2022).

37. Inamura, Y., Nakatani, T., Suzuki, J. & Otomo, T. Development Status of Software "Utsusemi" for Chopper Spectrometers at MLF, J-PARC. *J. Phys. Soc. Japan* **82**, SA031 (2013).

38. Azuah, R. T. *et al.* DAVE: A Comprehensive Software Suite for the Reduction, Visualization, and Analysis of Low Energy Neutron Spectroscopic Data. *J. Res. Natl. Inst. Stand. Technol.* **114**, 341-358 (2009).

39. Blaha, P. *et al.* WIEN2k: An APW+lo program for calculating the properties of solids. *J. Chem. Phys.* **152**, 074101 (2020).

40. Perdew, J. P., Burke, K. & Ernzerhof, M. Generalized Gradient Approximation Made Simple. *Phys. Rev. Lett.* **77**, 3865–3868 (1996).

41. White, S. R. Density matrix formulation for quantum renormalization groups. *Phys. Rev. Lett.* **69**, 2863–2866 (1992).

42. Schollwöck, U. The density-matrix renormalization group in the age of matrix product states. *Ann. Phys.* **326**, 96–192 (2011).

43. Alvarez, G. The density matrix renormalization group for strongly correlated electron systems: A generic implementation. *Comput. Phys. Commun.* **180**, 1572–1578 (2009).

44. Nocera, A. & Alvarez, G. Spectral functions with the density matrix renormalization group: Krylov-space approach for correction vectors. *Phys. Rev. E* **94**, 053308 (2016).

45. Nocera, A. & Alvarez, G. Root-*N* Krylov-space correction vectors for spectral functions with the density matrix renormalization group. *Phys. Rev. B* **106**, 205106 (2022).

## Acknowledgments

We thank D. I. Khomskii, I. Zaliznyak, A. M. Tsvelik, K. Nakajima, S. E. Nagler, J. Fernandez-Baca, Y. M. Qiu, and W. Yang for helpful discussions. AN acknowledges computational resources and services provided by Advanced Research Computing at the University of British Columbia. The AMATERAS and DNA experiments were performed under the auspices of the user program at the Materials and Life Science Experimental Facility of the J-PARC (Proposals #2020B0107 and #2022A0069). Work at Texas A&M University (XYL) was supported by the National Science Foundation through grant NSF-DMR-1807451. Work at UBC (XYL, MCA, MO, AN, KF, GS) was supported by the Natural Sciences and Engineering Research Council of Canada (NSERC), and through the Stewart Blusson Quantum Matter Institute by the Canada First Research Excellence Fund (CFREF).

## Author contributions

XYL and MO grew the single crystals and characterized them. XYL, NM, and MK performed neutron scattering experiments on AMATERAS at J-PARC, while XYL, MM, and HT performed neutron scattering experiments on DNA at J-PARC. XYL analyzed the neutron scattering data in consultation with MCA. AN carried out DMRG calculations, and KF carried out DFT calculations in consultation with GS. XYL and MCA wrote the paper with contributions from all the authors.

## Competing interests

The authors declare no competing interests.

## Additional information

**Supplementary Information** is available in the online version of the paper.

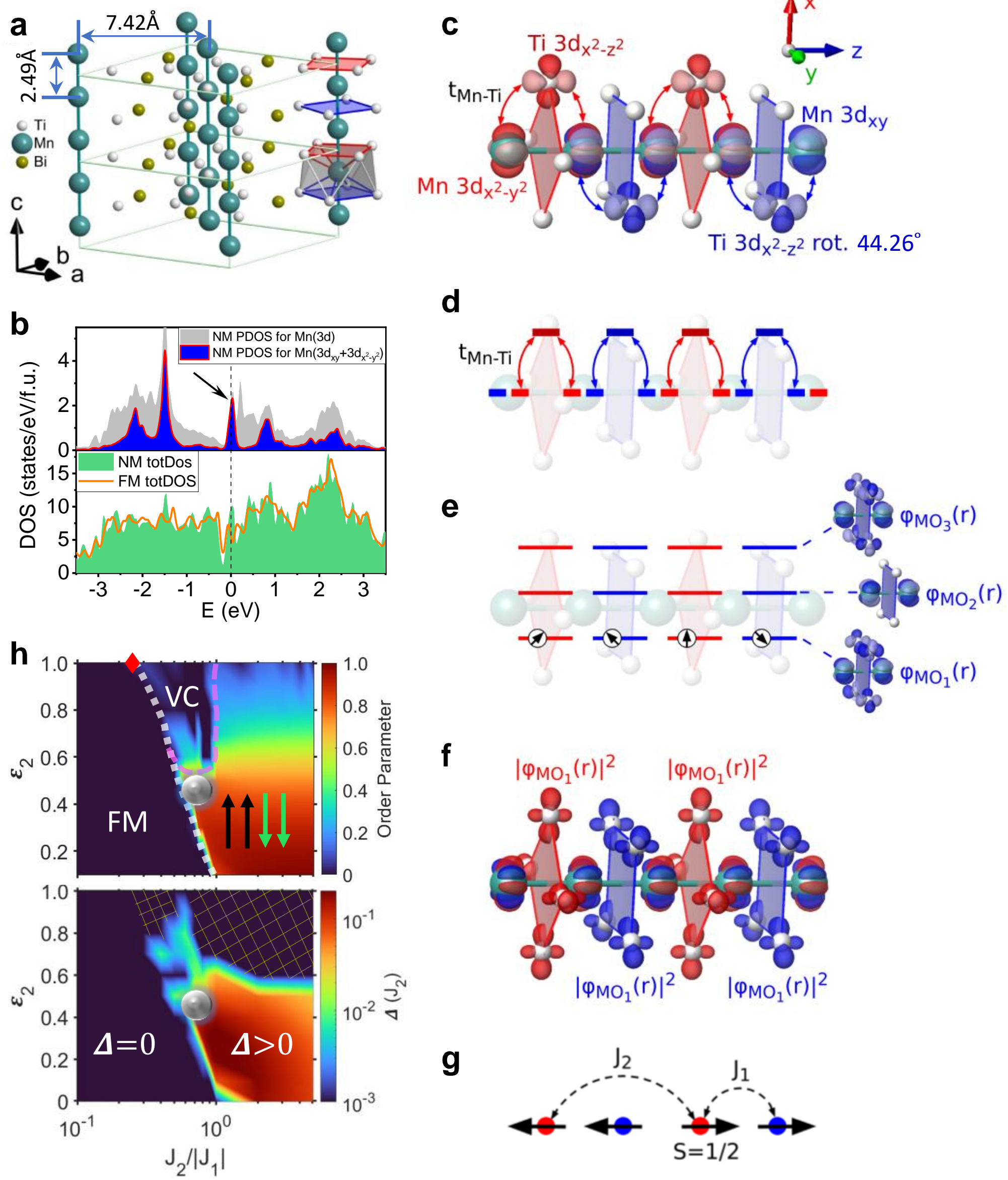

**Fig. 1 | Spin $S$ = 1/2 chains in $Ti_4MnBi_2$.** **a**, Mn chains in $Ti_4MnBi_2$ with square antiprismatic coordination from Ti squares, with relative rotations of 44.26°. **b**, (Top panel) Mn $d_{xy}$ and $d_{x^2-y^2}$ orbitals dominate the narrow peak at $E_F$ in the non-magnetic (NM) DFT Mn 3$d$ projected density of states (PDOS). (Bottom panel) The Mn $d_{xy}$ and $d_{x^2-y^2}$ orbitals are fully polarized in the FM DFT solution, leaving only the contribution from the itinerant states to the DOS at $E_F$. **c**, Hopping integral $t_{\text{Mn-Ti}}$ couples $d_{xy}$ and $d_{x^2-y^2}$ orbitals to Ti $d_{x^2-z^2}$ orbitals, whose on-site energy levels are shown in **d**. One of the four Ti $d_{x^2-z^2}$ orbitals per Ti square is shown. **e**, Molecular orbitals (MO) result from hybridization of atomic orbitals in (c). Unpaired electron occupies MO $\varphi_{MO1}(r)$, giving $S$ = 1/2 per MO. **f**, Spatial distribution of magnetic moments is proportional to $|\varphi_{MO1}(r)|^2$, and is centered on the Ti squares. **g**, Near neighbor interaction $J_1$ is FM, and next nearest neighbor interaction $J_2$ is AF. **h**, AF order parameter $O_{\uparrow\uparrow\downarrow\downarrow}$ (top) and magnetic gap $\boldsymbol{\Delta}$ (bottom) as functions of $J_1$-$J_2$ XXZ parameters $\boldsymbol{\alpha} = J_2/|J_1|$, $\boldsymbol{\varepsilon}_2$ at $T = 0$ (see text). $Ti_4MnBi_2$ is given by the gray circles. For continuity to the isotropic limit ($\boldsymbol{\varepsilon}_2$ = 1) the cross-hatched region is presumed to have a vanishingly small gap[13]. Details of DFT and DMRG calculations in SI 3, 4.

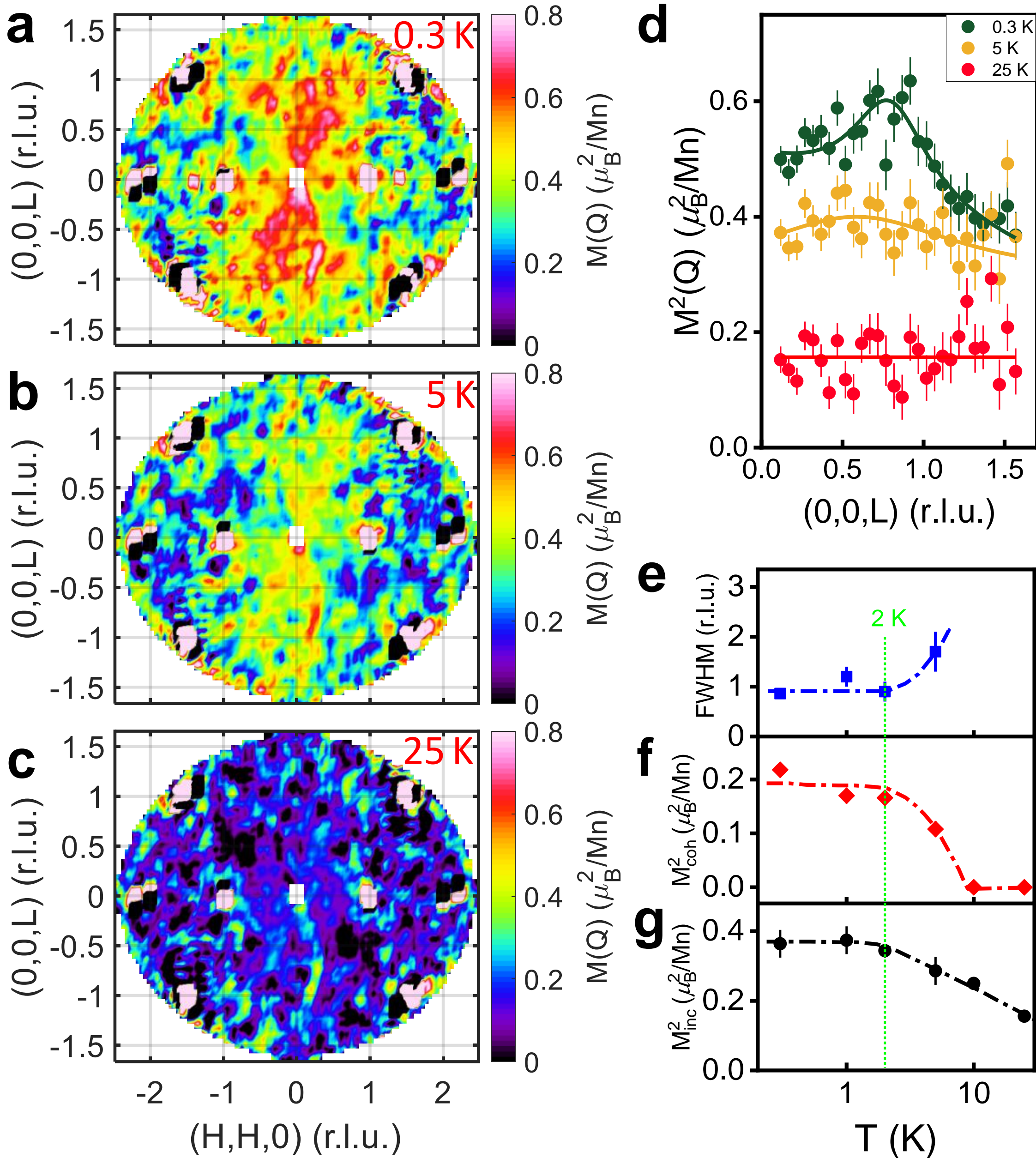

**Fig. 2 | Emerging magnetic correlations in $Ti_4MnBi_2$.** The static magnetic structure factor $M(\boldsymbol{Q})$ is shown at **a**, 0.3 K, **b**, 5 K, and **c**, 25 K for $\boldsymbol{Q}_{\mathbf{L}}$ parallel and $\boldsymbol{Q}_{\mathbf{HH}}$ perpendicular to the chain, obtained by integrating the scattered intensity between [-0.1, 0.1] meV, and averaging over ($H$, -$H$, 0) = [-0.5, 0.5] r.l.u. 100 K data used as a background. **d**, Static structure factor, $M(\boldsymbol{Q})$, obtained by averaging data in (a), (b), and (c), over ($H$, $H$, 0) = [-0.5, 0.5] r.l.u. A broad peak is found below ~ 10 K, centered at $\boldsymbol{Q}^*$ ~ (0, 0, 0.76(4)) r.l.u. and FWHM = 0.86(8) r.l.u for $T$ = 0.3 K. The peak is fitted at each temperature to a Lorentzian function with a sloping background, giving the temperature dependencies of the FWHM **(e)**, the Lorentzian intensity $M^2_{coh}$**(f)** and the incoherent scattering $M^2_{inc}$ (**g**). For the purposes of this experiment, scattering with energies less than the AMATERAS instrumental resolution of ~ 0.06 meV is effectively elastic. Experimental details in SI 1, 2.

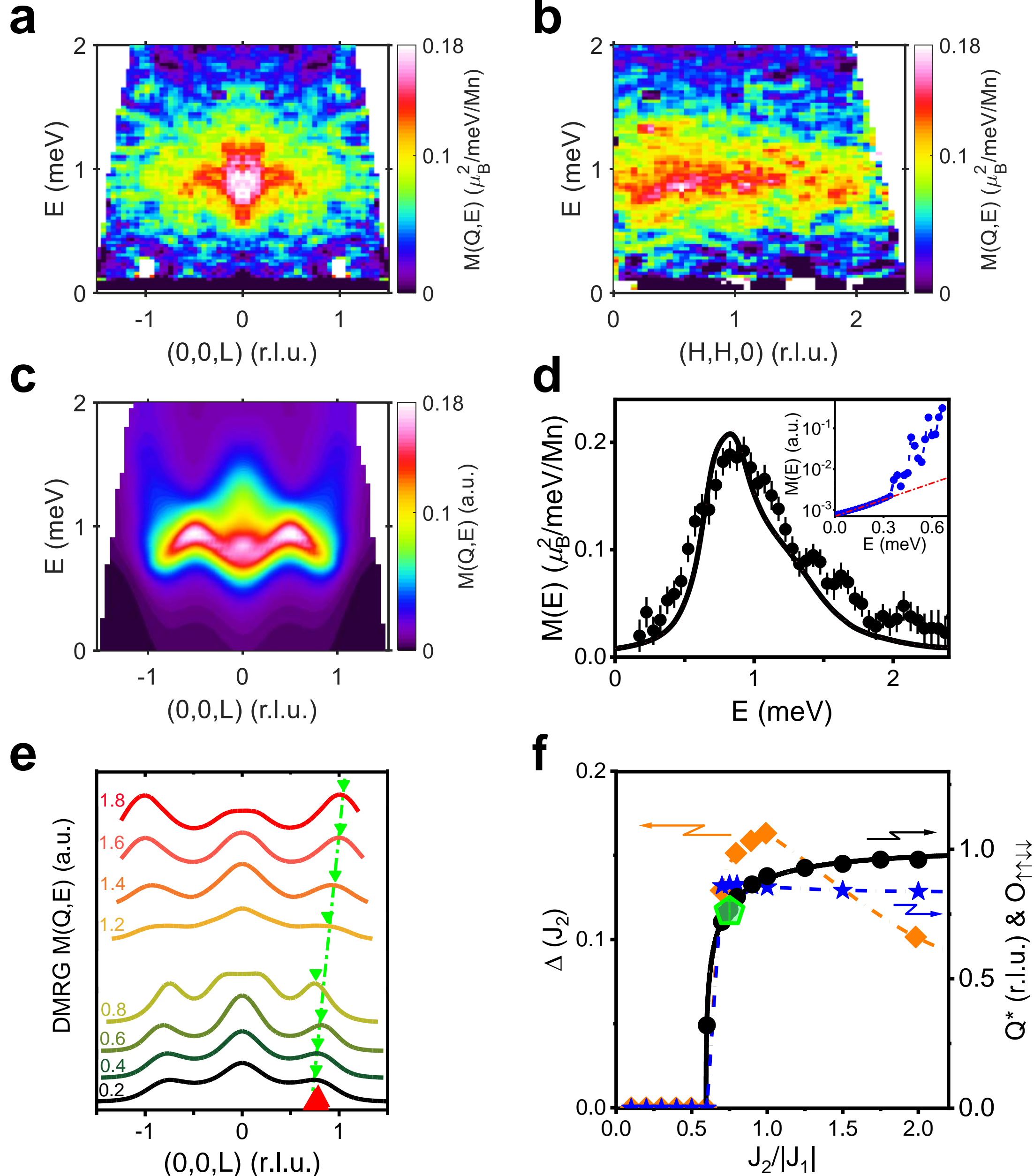

**Fig. 3 | Spinons and helical modes in $Ti_4MnBi_2$.** Magnetic dynamical structure factor $M(\boldsymbol{Q}, E)$ at 0.3 K **a**, for $\boldsymbol{Q}_{L} = (0, 0, L)$ summed over $\boldsymbol{Q}_{HH} = [0, 2]$ r.l.u, and **b**, for $\boldsymbol{Q}_{HH} = (H, H, 0)$ summed over $\boldsymbol{Q}_{L} = [-1, 1]$ r.l.u, and $(H, -H, 0)$ for [-0.5, 0.5] r.l.u. **c**, $M(\boldsymbol{Q}, E)$ from DMRG with $J_1$-$J_2$ XXZ parameters $\boldsymbol{\alpha} = J_2/|J_1| = 0.75$ and $\boldsymbol{\varepsilon}_2 = 0.425$ (SI 4.5). **d**, $M(\boldsymbol{Q}_{L} = 0, E)$ from INS (circles) and DMRG (solid line, broadened from the INS instrumental resolution of 0.06 meV to 0.13 meV). Inset: DMRG with high energy resolution 0.0024 meV shows gap for $E < 0.35$ meV. **e**, Energy cuts of the DMRG $M(\boldsymbol{Q}, E)$, with dispersing helimagnons (green triangles), and elastic peak $\boldsymbol{Q}^* = 0.76(4)$ r.l.u. (red triangle). **f**, DMRG calculated values of the gap $\boldsymbol{\Delta}$, the AF order parameter $O_{\uparrow\uparrow\downarrow\downarrow}$, and the helimagnon $\boldsymbol{Q}^*$ as functions of $\boldsymbol{\alpha} = J_2/|J_1|$, for fixed $\boldsymbol{\varepsilon}_2 = 0.425$. Black line is a guide for the eye. Green pentagon has $\boldsymbol{Q}^* = 0.76(4)$ r.l.u. Experimental details in SI 1, 2 and SI 4.5.

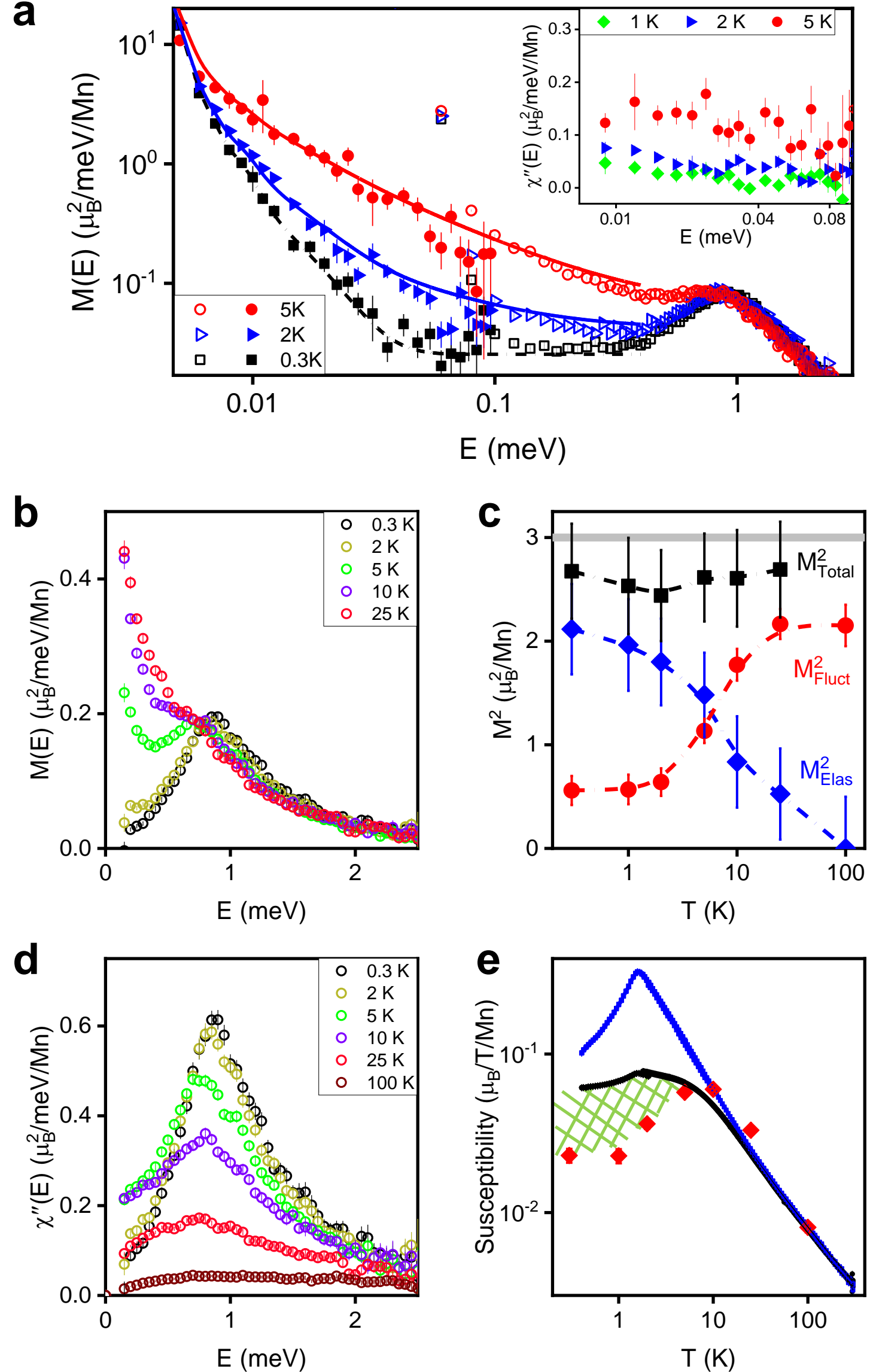

**Fig. 4 | Temperature dependence of the spin dynamics in $Ti_4MnBi_2$.** **a**, Powder averaged $M(E)$ measured at AMATERAS (open symbols) and DNA (solid symbols). Black dashed line is elastic peak convolved with DNA instrumental resolution, while blue and red lines indicate Bose factors at 2, and 5 K. Inset: $\chi''(E)$ where 0.3 K data are used as background. **b**, $M(\boldsymbol{Q}_{\mathrm{L}} = 0, E)$, averaged over (0, 0, $L$) = [-0.1, 0.1] r.l.u., ($H$, $H$, 0) = [0, 2] r.l.u., and ($H$, -$H$, 0) = [-0.5, 0.5] r.l.u. **c**, Elastic moment $M^2_{\mathrm{Elas}}(T)$ integrated $M(\boldsymbol{Q}, E)$ over energies [-0.1, 0.1] meV, and wave vectors where strongest magnetic diffuse scattering is found, i.e., (0, 0, $L$) = [0, 1] r.l.u., ($H$, $H$, 0) = [-0.5, 0.5] r.l.u., and ($H$, -$H$, 0) = [-0.5, 0.5] r.l.u. Fluctuating moment $M^2_{\mathrm{Fluct}}(T)$, integrated over energies [-2.4, -0.1] and [0.1, 2.4] meV, and wave vectors in the first BZ. $M^2_{\mathrm{Total}} = M^2_{\mathrm{Elas}} + M^2_{\mathrm{Fluct}}$ ~ 90% of $S$ = 1/2 moment (~ 3 $\mu_{\mathrm{B}}^2$/Mn, gray line). Details in SI 2.4. **d**, $E$-dependencies of $\chi''(\boldsymbol{Q}_{\mathrm{L}} = 0, E)$. **e**, The static susceptibility $\chi(T)$ determined (SI 2.5) from the Kramers-Kronig relation (red symbols) is compared to $M/H$ measured with $H$ = 100 Oe // $c$-axis (black points) and $H$ = 100 Oe ⊥ $c$-axis (blue points). Green cross-hatched region is the extrapolated contribution to $\chi(T)$ from the elastic scattering, excluded from the Kramers-Kronig analysis.